\documentclass[10pt,aps,twocolumn,showpacs,amsmath,amssymb,pre,superscriptaddress, floatfix, longbibliography]{revtex4-1}

\usepackage{graphicx}% Include figure files
\usepackage{dcolumn}% Align table columns on decimal point
\usepackage{bm}% bold math
\usepackage{amssymb}
\usepackage{mathtools}
\usepackage{multirow}
\usepackage{braket}

\usepackage{hyperref}
\usepackage{xcolor}

\begin{document}

% \title{Local symmetries and defect detection in static and periodically driven lattices}
% \title{Local symmetries and defect detection in periodically driven lattices}
\title{Exposing local symmetries in distorted driven lattices via time-averaged invariants}

\author{T. Wulf}
    \email{Thomas.Wulf@physnet.uni-hamburg.de}
    \affiliation{Zentrum f\"ur Optische Quantentechnologien, Universit\"at Hamburg, Luruper Chaussee 149, 22761 Hamburg, Germany}

\author{C. V. Morfonios}
    \affiliation{Zentrum f\"ur Optische Quantentechnologien, Universit\"at Hamburg, Luruper Chaussee 149, 22761 Hamburg, Germany}
    
\author{F. K. Diakonos}
  \affiliation{Department of Physics, University of Athens, GR-15771 Athens, Greece}

 \author{P. Schmelcher}
    \email{Peter.Schmelcher@physnet.uni-hamburg.de}
    \affiliation{Zentrum f\"ur Optische Quantentechnologien, Universit\"at Hamburg, Luruper Chaussee 149, 22761 Hamburg, Germany}
    \affiliation{Hamburg Centre for Ultrafast Imaging, Universit\"at Hamburg, Luruper Chaussee 149, 22761 Hamburg, Germany} 

\date{\today}

\pacs{05.45.Mt,05.60.Gg,05.45.Pq}

\begin{abstract}
Time-averaged two-point currents are derived and shown to be spatially invariant within domains of local translation or inversion symmetry for arbitrary time-periodic quantum systems in one dimension.
% By calculating these currents for static and driven lattices, local deviations from a given spatial symmetry can be extracted reliably from stationary or quasistationary wavefunctions. 
These currents are shown to provide a valuable tool for detecting deformations of a spatial symmetry in static and driven lattices.
In the static case the invariance of the two-point currents is related to the presence of time-reversal invariance and/or probability current conservation.
The obtained insights into the wavefunctions are further exploited for a symmetry-based convergence check which is applicable for globally broken but locally retained potential symmetries.
% Our findings reveal a systematic encoding of static and dynamical local symmetries of general potentials into their eigenstates, and is particularly relevant for strongly driven systems where local order is obscured in the wavefunction by its dynamics.
\end{abstract}

\maketitle

\section{Introduction}

The analysis of the symmetries of a given physical system is often one of the cornerstones of its thorough theoretical description. 
A paradigmatic example is Bloch's theorem which governs the form of the solutions of the Schr\"odinger equation in the presence of a discrete translational symmetry \cite{bloch_uber_1929} and is applied routinely in the investigation of band structure in crystals \cite{Introduction_2001}. 
Analogously, systems with inversion symmetry yield energy eigenstates of even or odd parity, with profound consequences for their description such as resulting selection rules for transitions among states \cite{zettili_quantum}.
The imprints of translation and inversion symmetry on wavefunctions have also been formulated for driven quantum systems in the form of generalized Bloch states in periodically driven lattices \cite{Miyazaki1994_PRE_50_910_Quasienergy-bandStructure} or eigenstates of spatial parity combined with a temporal shift \cite{Breuer1988_ZFPDA_8_349_RoleAvoided,Grossmann1991_PRL_67_516_CoherentDestruction} or inversion \cite{Lehmann2003_JCP_118_3283_RectificationLaser-induced}.
Even more, and in particular for the driven lattices it was realized how the analysis of the systems spatiotemporal symmetries allows for the decisive insights into the transport properties 
of a given setup \cite{Schanz_classical_2001, schanz_directed_2005, denisov_periodically_2007, wulf_symmetries_2014}.

All of the symmetry implications described above, rely on the assumption of a \textit{globally} valid symmetry which, however, can only be seen as an approximation: 
Firstly, it contrasts the finiteness of any physical system and, secondly, it neglects the occurrence of any type of defects, i.\,e. of local deviations from the underlying symmetry. 
In other words, spatial symmetries hold strictly only \textit{locally} rather than globally in real systems.
There are in fact many scenarios where a system is naturally made up of different domains in which distinct spatial symmetries hold locally, striking examples being quasicrystals \cite{shechtman_metallic_1984, kraemer_embedding_2013, morfonios_local_2014} or large molecules \cite{pascal_concise_2001, domagala_optimal_2008}. 
The quest of finding symmetry remnants in stationary states of systems with globally broken but locally retained symmetries recently led to a generic treatment of one-dimensional (1D) local symmetries in terms of invariant two-point currents \cite{kalozoumis_invariants_2014}.
These symmetry-induced spatial invariants were shown to enable a mapping of wave amplitudes of scattering states \cite{kalozoumis_invariants_2014} or basis functions \cite{zambetakis_invariant_2015} between symmetry-related points within any local symmetry (LS) domain.
They have also been used to classify perfect transmission resonances \cite{kalozoumis_local_2013} and as an order parameter for spontaneous symmetry breaking in parity-time symmetric scattering \cite{kalozoumis_systematic_2014} and
apply to a large variety of systems described by a spatial Helmholtz equation as e.\,g. in classical optics \cite{peng_symmetry-induced_2002}, quantum scattering \cite{kalozoumis_local_2013}, or even acoustic wave propagation \cite{hladky-hennion_acoustic_2013, martinez-gutierrez_transverse_2014, theocharis_limits_2014} where it was readily corroborated by experiments \cite{kalozoumis_invariant_2015}.
As the invariant two-point currents are calculated from the stationary states, the LS formalism was so far --by construction-- restricted to time-independent setups, thus making 
its generalization to time-dependent systems an intriguing yet challenging task.
Here, the reflection of LS in the spatial structure of nonequilibrium states could provide valuable insight into experiments with driven quantum systems as performed, e.\,g., with cold atoms in shaken optical lattices \cite{salger_directed_2009, salger_tuning_2013} or with radiated semiconductors \cite{glazov_high_2014, olbrich_classical_2011}. 
For both static and driven potentials, an intriguing prospect would then be to utilize the correspondence between LS and induced wavefunction invariants for the \textit{detection} of local order in extended systems.

In the present work, we derive a generalization of LS-induced invariants to periodically time-dependent quantum systems in the framework of Floquet theory.
Specifically, we show that the time-average of the two-point current of any Floquet mode behaves as its static counterpart, meaning that it is spatially constant in arbitrary domains of translation or inversion symmetry. 
% This dynamical two-point current thereby also extends the established analysis of global symmetries in driven systems \cite{Schanz_classical_2001, schanz_directed_2005, denisov_periodically_2007, wulf_symmetries_2014} to local ones.
By applying the formalism to static and driven lattices, we show how the locations of defects in an otherwise symmetric potential are encoded in the stationary or quasistationary states of the system and can be extracted by use of the LS invariants. 
Furthermore, we propose a convergence measure for numerically obtained solutions of globally non symmetric potentials which distinguishes between errors caused by finite spatial resolution and by basis truncation.
Finally, a physical perspective is given to the two-point currents by linking their spatial constancy to the general properties of probability current conservation (PCC) and time reversal invariance (TRI) in terms of amplitude transfer matrices.

The paper is organized as follows.
In Sec.\,\ref{sec:loc_sym_driven} we derive the dynamical two-point current for arbitrary locally translation- or inversion-symmetric periodically driven systems.
In Sec.\,\ref{sec:loc_sym_lattices} the currents are calculated for static and driven superlattices of Gaussian barriers, demonstrating the extraction of symmetry domains and defect locations from the LS-induced invariant current.
Finally, a powerful convergence measure for numerical wavefunction computations is introduced.
In Sec.\,\ref{sec:transfer_matrices} the two-point current invariance is related to PCC and TRI within a transfer matrix approach.
Sec.\,\ref{sec:conclusion} concludes our findings.

\section{Symmetry-induced invariants in periodically driven systems}
\label{sec:loc_sym_driven}

\begin{figure}[t!]
\centering
\includegraphics[width=.9\columnwidth]{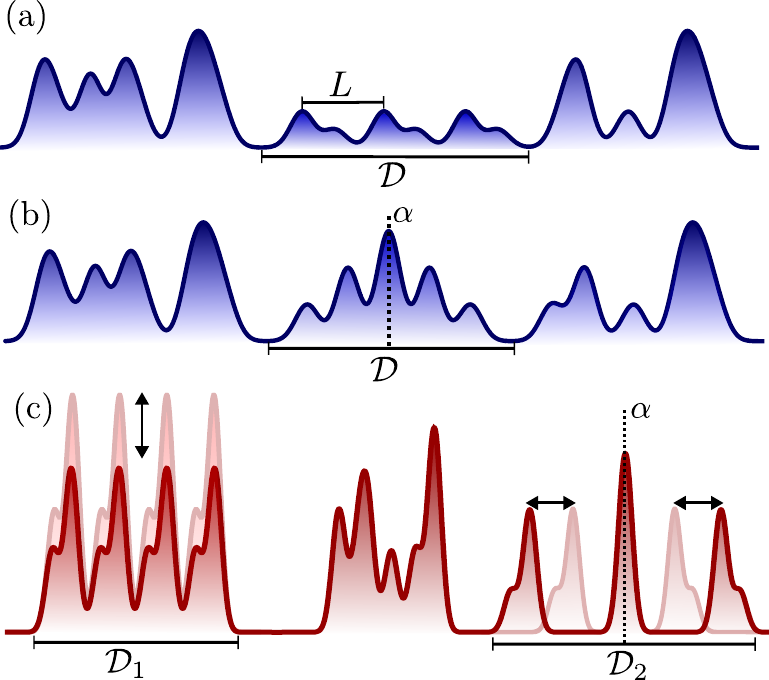}
\caption{\label{fig:sketch} 
Illustrative sketches of static potential landscapes possessing a local symmetry for $x,\bar{x} \in \mathcal{D}$ under (a) translation by $L$ and (b) inversion through position $\alpha$. (c) 
Sketch of a time-dependent potential with $V(x,t)=V(\bar{x},t)$ within localized domains of translation ($\mathcal{D}_1$) and inversion ($\mathcal{D}_2$) symmetry.}
\end{figure}

Let us start by defining the notion of local symmetries in one spatial dimension and the induced invariant quantities for a static potential (see \cite{kalozoumis_invariants_2014}).
For a quantum particle of energy $E$ obeying the 1D stationary Schr\"odinger equation (setting $\hbar=m=1$)
\begin{equation}
 \Psi''(x) + 2 [E - V(x)]\Psi(x) = 0,
 \label{eq:schrodinger}
\end{equation}
it is well established how global symmetries of a real time-independent potential $V(x)$ provide information on symmetry properties of a stationary state $\Psi(x)$. 
For example, Bloch's theorem dictates that if the potential possesses a discrete translation symmetry, i.\,e. $V(x)=V(x+L)$ for some fixed length $L$, then the energy eigenstates obey $\Psi(x+L)=e^{i\kappa L} \Psi(x)$ with $\kappa$ being the real quasimomentum.
Analogously, for a potential with inversion symmetry, $V(x)=V(-x)$, the eigenstates can be classified into even and odd parity states: $\Psi(-x)=\pm\Psi(x)$.
However, the application of Bloch's theorem or the parity theorem both require the potential and the boundary conditions to respect the symmetry globally and thus, strictly speaking, cannot be applied in cases where it is symmetric only within some finite domain $\mathcal{D}\subset \mathbb{R}$. 
Illustrative examples of potentials obeying translation or parity symmetry only locally, i.\,e. only for $x\in \mathcal{D}$, are shown in Figs.\,\ref{fig:sketch} (a) and (b), respectively.

% Such a static LS gives rise to `two-point currents' which are spatially constant within a LS domain and in this sense reflect the information of the potential symmetry directly in the wavefunction.
Specifically, let us assume that $V(x)=V(\bar{x})$ for all $x,\bar{x} \in \mathcal{D}$ and with $\bar{x}= x+L$ (translation by $L$) or $\bar{x}= -x+2\alpha$ (inversion through $\alpha$).
It can then be shown \cite{kalozoumis_invariants_2014} that, for a stationary state $\Psi(x,t) = \Psi(x)e^{-iEt}$, the two-point current
\begin{equation} \label{eq:Qstatic}
{{Q}}_\Psi (x,\bar x,t) = \frac{1}{2i}\left[\sigma \Psi^*(x,t)\Psi'(\bar{x},t) - \Psi(\bar{x},t)\Psi^{*\prime}(x,t)\right] 
\end{equation}
is spatially constant within the LS domain at any instance $t$, i.\,e. ${{Q}}_\Psi (x,\bar x,t)' = 0$ for $x,\bar{x} \in \mathcal{D}$, where $\sigma$ distinguishes between the cases of translation ($\sigma=+1$) and inversion symmetry ($\sigma=-1$). Furthermore,
we use the shorthand notation $\Psi'(\bar{x}) \equiv \partial \Psi( x) / \partial x |_{x=\bar{x}}$ throughout the paper. 
Hence, any static LS gives rise to `two-point currents' which are spatially constant within a LS domain and in this sense reflect the information of the potential symmetry directly in the wavefunction.
For $\sigma = +1$, the limit of ${{Q}}_\Psi$ for zero translation, $\bar{x}=x$, yields the usual probability current $J(x,t) = {{Q}}(x,x;t)$, whose global spatial constancy for any 1D (real) potential is thus recovered.
There is also a complementary two-point invariant quantity, which we here denote $Q^\text{c}_\Psi (x,\bar x, t)$, obtained simply by replacing $\Psi^*$ by $\Psi$ in Eq.\,(\ref{eq:Qstatic}). 
Note that, while ${{Q}}_\Psi$ is time-independent for a static potential, $Q^\text{c}_\Psi$ varies in time through the phase factor $e^{-2iEt}$, thus having only a time-independent modulus. 
We concentrate here on ${{Q}}_\Psi$, in terms of which an invariant for time-periodic systems is derived in the following.

%%%%%%%%%%%%%%%%%%%%%%%%%%%%%%%%%%%%%%%%%%%%%%%%%
%%%%%%%%%%%%%%%%%%%%%%%%%%%%%%%%%%%%%%%%%%%%%%%%%

Let us now turn to a time-dependent potential $V(x,t)$ with the time-dependent Schr\"{o}dinger equation (TDSE)
\begin{equation}
 i \frac{\partial}{\partial t} \Psi(x,t) = -\frac{1}{2}\frac{\partial^2}{\partial x^2}\Psi(x,t)+ V(x,t)\Psi(x,t)
 \label{TDSE}
\end{equation}
In the following
we will show that if the potential varies periodically in time with frequency $\omega$,
\begin{equation} \label{eq:Vperiodic}
 V(x,t) = V(x,t+T), ~~ T=2\pi / \omega,
\end{equation}
then there exist \textit{dynamical} invariants which are spatially constant in LS domains.
Considering a time-periodic potential, Floquet's theorem ensures that the solutions of Eq.\,(\ref{TDSE}) are
of the form % $\Psi_{{{m}}}(x,t) = e^{-i\epsilon_{{{m}}}t} \Phi_{{{m}}}(x,t)$ 
\begin{equation} \label{eq:Floquet}
 \Psi_{{}}(x,t) = e^{-i\epsilon_{{}}t} \Phi_{{}}(x,t)
\end{equation}
with the real quasienergy $\epsilon_{{}} \in [-\omega/2, +\omega/2]$ and with $\Phi_{{}}(x,t) = \Phi_{{}}(x,t+T)$. 
That is, the solutions of the TDSE with a time-periodic potential are also periodic apart from an oscillating phase factor. 
The states $\Psi_{{}}(x,t)$, commonly referred to as Floquet modes in analogy to Bloch modes of a spatially periodic potential, will now be used to construct a dynamically invariant current for the periodically driven system. 
To this end, we assume the potential retains its local spatial symmetry during the driving, that is, $V(x,t) = V(\bar{x},t)$ for $x,\bar{x} \in \mathcal{D}$ and for any $t$
(see Fig.\ref{fig:sketch} (c) for a sketch of a periodically time-dependent potential possessing different domains of local translation and inversion symmetry). 
By multiplying Eq. (\ref{TDSE}) evaluated at position $x$ by $\Psi^*_{{}}(\bar{x},t)$ and subtracting its complex conjugate evaluated at position $\bar{x}$ and multiplied by $\Psi_{{}}(x,t)$, we obtain a nonlocal version of the continuity equation,
\begin{align}
  \frac{\partial}{\partial x}{{Q}}_{\Psi}(x,\bar{x};t) + \frac{\partial}{\partial t}[ \Psi_{{}}(x,t)\Psi^*_{{}}(\bar{x},t)] = 0,
 \label{eq:continuity}
\end{align}
for the two-point current defined in Eq.\,(\ref{eq:Qstatic}).
Clearly, since the associated state $\Psi_{{}}(x,t)$ is not stationary, the current ${{Q}}_{\Psi}$ will no longer be spatially invariant. 
Instead, we can exploit Floquet's theorem by plugging in the ansatz of Eq.\,(\ref{eq:Floquet}) into Eq. (\ref{eq:continuity}), and subsequently integrate over one period of the driving, which yields
\begin{align}
 \frac{\partial}{\partial x} \int_0^T dt \, {{Q}}_{\Phi}(x,\bar{x};t) + \left[ \Phi_{{}}(x,t)\Phi^*_{{}}(\bar{x},t)\right]^T_0 = 0
 \label{continuity2}
\end{align}
noting that ${{Q}}_{\Psi} = {{Q}}_{\Phi}$.
Now, since $\Phi_{{}}(x,0)=\Phi_{{}}(x,T)$, the second term in Eq.\, (\ref{continuity2}) vanishes and we obtain the desired locally invariant current for periodically driven systems,
\begin{equation}
{\bar{Q}}_{\Phi}(x,\bar{x}) \equiv \frac{1}{T} \int_0^T dt \, {{Q}}_{\Phi}(x,\bar{x};t)
\label{eq:Qdynamical}
\end{equation}
obeying $\bar{Q}_{\Phi}' = 0$ for $x,\bar{x} \in \mathcal{D}$. %(with the Floquet mode superscript suppressed for notational simplicity).

We have thus found a quantity $\bar{Q}_{\Phi}$ which, remarkably enough, identifies through its spatial constancy the instantaneous local symmetries of any time-periodic potential, however strong the driving or the spatiotemporal modulation of the modes $\Phi_{{}}$ (and thereby of ${{Q}}_{\Phi}$) may be.
Note that there is no such dynamical equivalent for the complementary static invariant $Q^\text{c}$.
In the coincidence limit $\bar{x}=x$, the condition $\bar{Q}_{\Phi}'=0$ yields the global spatial constancy of the period-averaged probability current $\bar{J}(x) = \bar{Q}_{\Phi}(x,x)$ which holds for any Floquet mode for arbitrary 1D real time-periodic potentials.

\section{Local symmetries in static and driven lattices}
\label{sec:loc_sym_lattices}

In the remainder of the manuscript we will show how the LS formalism can be applied to static and periodically driven lattices. In particular we argue how 
local distortion from a translational symmetry can be be extracted from the systems stationary or quasistationary states (since the analysis for inversion symmetry, and local deviations thereof,
works in precisely the same manner, we refrain from showing it).
Even though the LS formalism above applies to any static or periodically driven potential $V(x,t)$ possessing one or several domains of LS, we will now make our discussion more concrete
and focus on potentials of the form
\begin{equation}  \label{eq:lattice_pot}
 V(x,t)=\sum_{n=0}^{N-1} V_n(x-X_n,t),
\end{equation}
describing $N$ scatterers localized at positions $X_n\equiv nL$ for some spacing $L$, where the functions $V_n(x,t)$ are assumed to have compact support of width $w<L$ (i.\,e. $V_n(x)=0$ for $|x|>w/2$) so that the scatterers do not overlap at any instance in time.
A breaking of the global translational symmetry into domains of local symmetry is achieved by first setting $V_n=V$ and then altering one or several of the $V_n$, thus creating local impurities in the lattice.
We will here show how any local symmetry---or deviations from it---can be extracted reliably by means of the two-point currents for static or driven systems.
As a concrete example, we consider scatterer potentials of Gaussian profile,
\begin{equation}
 V_n(x,t) = \Lambda_n \exp\left[-\frac{(x - d(t))^2}{\Delta^2} \right],
 \label{eq:gaussian}
\end{equation}
with strengths $\Lambda_n$ and of common widths $\Delta$, chosen such that the $V_n(x,t)$ practically fall off to zero within $|x|<w/2$ for some $w$.
For the time-periodic modulation we opt for a lateral shaking of the barriers by a periodic driving law $d(t)$ 
(equivalently, one could e.g. drive the potential strengths periodically).
Any such case can be efficiently treated numerically (see below) and at the same time provides a relatively realistic interaction profile (compared to, e.\,g., piecewise constant potentials).

Being time-independent functions, $Q$ and $\bar{Q}$ can both be
employed to explore the symmetries in a specific setup being driven (using
$\bar{Q}$ when $d(t)\neq 0$) or static (using $Q$ for $d(t)=0$), in a completely equivalent manner.
% Being time-independent functions, ${{Q}}$ and $\bar{Q}$ are completely equivalent in the symmetry analysis of the considered systems, meaning that any symmetry extraction in a static setup (using ${{Q}}$) will work analogously for the same setup driven by $d(t)$ (and using $\bar{Q}$ for the symmetry analysis).
Therefore, in order to introduce the concept of symmetry extraction in a clear manner, we will start with the simpler scenario of a static lattice containing a single impurity; then subsequently increase the setups complexity and ultimately, arrive at the periodically driven case. 

\subsection{Local breaking of symmetries in distorted lattices}
\label{sec:static_lattice}

It goes without saying that potentials of the form of Eqs.\, (\ref{eq:lattice_pot})  and (\ref{eq:gaussian}) quite generically require a numerical solution of the corresponding Schr\"odinger equation whereas 
analytical solutions may be found only in very few limiting cases. 
In fact, one of the most minimalistic setups captured by Eq.\, (\ref{eq:gaussian}) is a chain of static delta potentials obtained by setting $d(t)=0$ and by letting $V_n \rightarrow \infty$ and simultaneously 
$\Delta \rightarrow 0$. For this very simple case, we present the analytically obtained invariant current $Q(x,\bar x)$, for $\bar x= x+L$ and for a single impurity in appendix \ref{A1}. 
Indeed we find that ${Q}'(x)=0$, except for a jump in ${{Q}}(x)$ precisely at the defect position for any energy of the incoming wave. 
Hence, by calculating ${{Q}}(x)$ for a given symmetry transform, one can reconstruct the position of the defect.
Admittedly, in this very special case of point scatterers, a defect will most probably be revealed by the wavefunction itself at any energy through a different kink at that position.
For a general potential, however, structural irregularities may be far from obvious by the profile of any given stationary state and we show in the following
how the two-point currents can indeed be used to deduce symmetries of the potential and their local breaking from arbitrary stationary waves.

To demonstrate this, we employ a static ($d(t) = 0$ in Eq.\,(\ref{eq:gaussian})\,) array of $N$ Gaussian barriers.
The potential is centered at $x = 0$, with periodic boundary conditions imposed at $x=\pm R/2$ as a natural choice for lattice systems (recall that the invariance of ${{Q}}$ depends only on the potential and applies for arbitrary boundary conditions).
In this sense, we will address the LS structure within a supercell of size ${{R}}$ in an infinitely extended periodic superlattice.
The energy eigenstates of the Hamiltonian $H = -\frac{\partial^2}{2\partial x^2} + V(x)$ are found numerically by diagonalizing its matrix representation $H_{\mu \nu}$ in a plane wave basis $\ket{\mu}$ respecting the periodic boundary conditions with $\braket{x | \mu } = \exp(i2\pi\mu x /{{R}})$
for $\mu=-k_{\text{max}},-k_{\text{max}} + 1,...,+k_{\text{max}}$, where $k_{\text{max}}$ is chosen large enough to ensure convergence.
The matrix elements $H_{\mu \nu} \equiv \braket{\mu|H|\nu}$ can then be evaluated analytically to
\begin{equation}
 H_{\mu \nu}=\frac{2 \pi^2}{{{R}}^2}\mu^2 \delta_{\mu \nu} + \frac{\sqrt{\pi}}{{{R}}} \sum_{n=1}^N W_{\mu\nu}^{(n)}
\end{equation}
with $W_{\mu\nu}^{(n)} \equiv \Lambda_n \Delta \exp\left[\frac{(\nu-\mu)\pi}{{{R}}^2}(2iX_n {{R}} -\Delta^2\pi (\nu-\mu) \right]$
containing the information on the potential barriers, where it was assumed that the potential has fallen off to zero at the supercell boundaries, $V(|x|=R/2) \sim 0$.

%%%%%%%%%%%%%%%%%%%%%%%%%%%%%%%%%%%%%%%%%%%%%%%%%
\begin{figure}[t!]
\centering
\includegraphics[width=1\columnwidth]{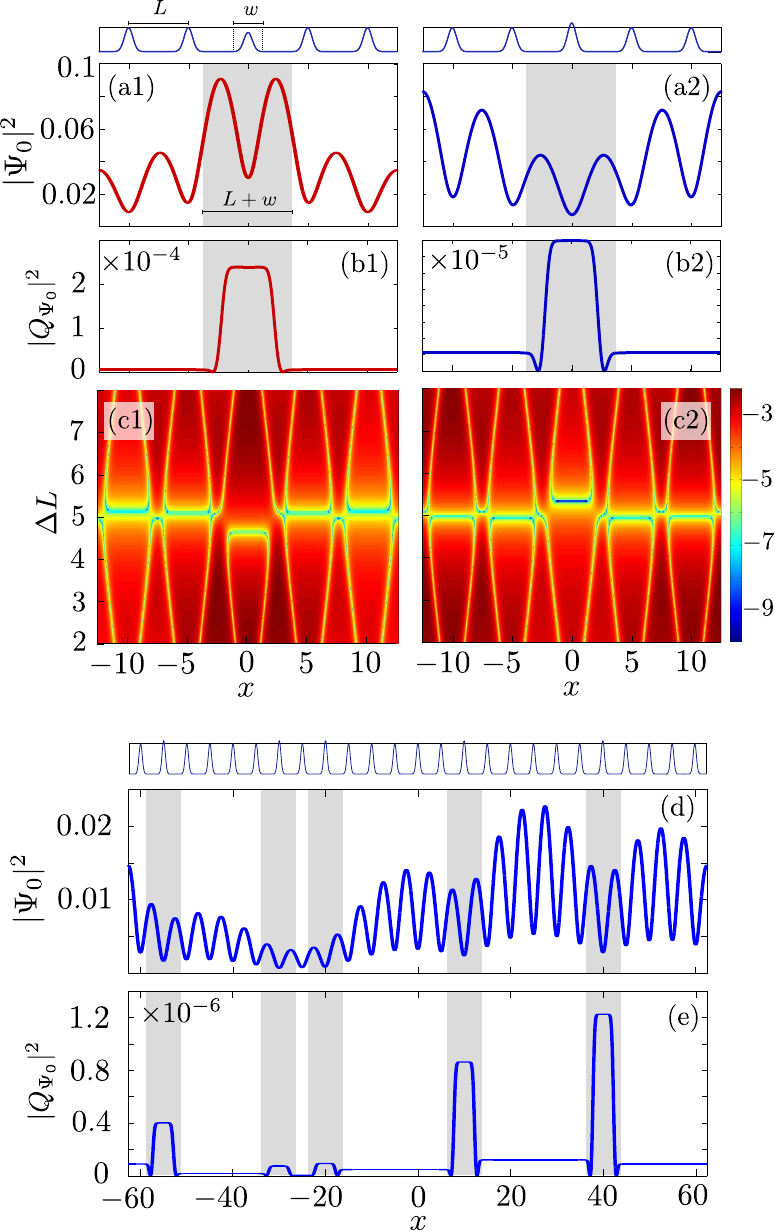}
\caption{\label{fig:static} 
(a1) Ground state probability densities $|\Psi_0|^2$ with corresponding (b1) nonlocal current $|{{Q}}_{\Psi_0}|^2$ (for translation $\bar{x} = x + L$ with $L=5$) as a function of $x$ and (c1) $\log_{10}[|{{Q}}_{\Psi_0}^{\Delta L}|^2]$ for varying $x$ and shift $\Delta L$ (see text), for a $N=5$-barrier superlattice with $V_n=1$ for $n\neq 0$ and $V_0=0.8$. 
The supercell potential is shown in arbitrary units in the top panel.
(a2,b2,c3) Same as above but for $V_0=1.2$.
(d) Density $|\Psi_0(x)|^2$ and (e) current $|{{Q}}_{\Psi_0}|^2$ for a lattice containing $N=25$ barriers in the supercell with five defects with $V_i = 1.1$ at positions $X_i=-55,-30,-20,10,x=40$. 
In all cases $|{{Q}}(x-L/2)|^2$ is shown so that the deviations from ${{Q}}_{\Psi_0}'=0$ are centered around the defect locations).}
\end{figure}
%%%%%%%%%%%%%%%%%%%%%%%%%%%%%%%%%%%%%%%%%%%%%%%%%

We now consider a setup with $N$ barriers of common strength $\Lambda$ in the supercell, and then implement defects by selectively setting $\Lambda_n \neq \Lambda$ for some $n$'s.
Then the supercell potential is periodic with period $L$ between the defect barriers.
With the given underlying translational symmetry $V(x)=V(x+L)$, we can now calculate ${{Q}}_\Psi$ which will thus be constant between the defects for a stationary state $\Psi$.
Specifically, ${{Q}}_\Psi'(x)$ may only deviate from zero if $x$ and/or $\bar x$ are within the interval where the defect barrier is nonzero, i.e. within the interval $(X_n-w/2,X_n+w/2)$.  
Hence, for the case of $\bar x=x+L$, we expect ${{Q}}_\Psi'(x)\neq 0$ 
for $x \in (X_n-w/2-L,X_n+w/2)$, that is, over an interval of size $L+w$ centered at $X_n - L/2$ for each defect $n$.
Hence, from these deviations the positions of the defects can be extracted.
This is exemplified for $N=5$ barriers (and thus $R=5\times L$) with an attractive or repulsive defect on the central barrier $n=0$ (that is, for $\Lambda_0 < \Lambda$ or $\Lambda_0 > \Lambda$, respectively).
The densities of the ground states, $\Psi_0$, are shown for both cases in Fig.\,\ref{fig:static}\,(a1,a2) and the resulting currents ${{Q}}(x)$ are shown in Fig.\,\ref{fig:static}\,(b1,b2).
In both cases, ${{Q}}_{\Psi_0}(x)$ clearly indicates the position of the defect barrier through the deviation from its constant value in the surrounding translationally symmetric region
(the same holds separately for the real and imaginary parts of ${{Q}}_{\Psi_0}(x)$; not shown).

Notably, in order to apply the above procedure, we have assumed the knowledge of the translational symmetry of the unperturbed lattice. 
Inversely, one could also ask whether a LS of the potential can in fact be deduced from an arbitrary stationary wavefunction $\Psi(x)$.
To answer this, we calculate ${{Q}}_{\Psi}^{\Delta L}(x) \equiv {{Q}}_{\Psi}(x,x+\Delta L)$ again for the ground state $\Psi_0$ with a varying shift parameter $\Delta L$ (the same could be done for inversion symmetry by varying the point of inversion $\alpha$).
Then ${{Q}}_{\Psi_0}^{\Delta L}$ generally changes in $x$, but becomes spatially constant whenever $\Delta L$ coincides with the period $L$ of a translational symmetry of the potential in some domain.
This is shown in Fig.\,\ref{fig:static}\,(c1,c2) for the same setup as above, which reveals that ${{Q}}_{\Psi_0}$ is indeed spatially constant away from the central defect only when the parametric shift coincides with the actual lattice spacing ($\Delta L = L = 5.0$). 
Hence, the $(x,\Delta L)$-profile of ${{Q}}_{\Psi}$ immediately reveals the lattice spacing of the underlying lattice potential as well as the defect location.

We complete this section by computing the ground state and the corresponding ${{Q}}$ for a larger and structurally more complex lattice (supercell) consisting of $N=25$ barriers with five 
arbitrarily chosen defect barriers; see Fig. \ref{fig:static} (d) and \ref{fig:static} (e), respectively. 
Although the strength of the defects is chosen to be small (only 10\% higher than the unperturbed barriers and thus hardly visible in the potential plot atop Fig.\ref{fig:static}\, (d)),
the two-point current reveals all defect positions reliably through its nonzero slope.
Note here that the constant ${{Q}}$-values between the defects are in general different;
for the example in Fig.\,\ref{fig:static}\,(a--c) with a single central defect they are equal due to the (additional) global inversion symmetry of the supercell.

\subsection{Defect detection through dynamical nonlocal invariants}
\label{sec:driven_lattice}

It was demonstrated above how the two-point current ${{Q}}_{\Psi}$ of any energy eigenstate enables the extraction of the local symmetries of a static potential.
We emphasize that the very same analysis applies to periodically driven potentials simply by replacing ${{Q}}_{\Psi}$ by its dynamical counterpart ${\bar {Q}}_{\Psi}$ for a \textit{quasienergy} eigenstate (Floquet mode) as defined in Eq.\,(\ref{eq:Qdynamical}).
In this subsection we underline the power of the dynamical invariant in revealing the spatial structure of a driven potential.
Indeed, one might argue that the wavefunction -especially in the ground state- of a static lattice potential often reveals defect positions directly since it will tend to be more (less) localized around a defect barrier which is lower (higher) than the surrounding ones.
This simple argument, however, is not reliable and typically not applicable to excited states. 
In particular, if the system is driven constantly out of equilibrium, even the energetically lowest (in a time-averaged sense) state may not reveal any obvious relations to defect positions.
For strongly driven systems it thus becomes even more challenging to deduce potential regularities from an arbitrary eigenstate.
This is achieved using the derived dynamical two-point invariant $\bar {Q}$, as we will showcase in the following.

We consider again the lattice potential defined by Eqs.\,(\ref{eq:lattice_pot}) and (\ref{eq:gaussian}), now with a periodic driving law
\begin{equation}
 d(t)=A\cos(\omega t),
 \label{eq:driving}
\end{equation}
that is, for a lateral shaking of the barriers.
Note that, by means of appropriate unitary transformations, the corresponding time-dependent Hamiltonian $H(t)$ can be brought in the form of a static lattice plus an oscillating force term. 
Thus, a potential of the form as given in Eq. (\ref{eq:gaussian}) may serve as a model potential for many cases of experimental interest, such as cold atoms in optical lattices where the lattice is actually laterally shaken (see e.g. \cite{salger_directed_2009}) or radiated semiconductors where an oscillating force term is added to a static lattice (e.g. \cite{glazov_high_2014}).

Numerically, the Floquet modes entering the ${\bar {Q}}$ in Eq.\,(\ref{eq:Qdynamical}) can be calculated by exploiting that the Floquet modes $\Psi(x,t)$ are eigenstates of the time evolution operator $U(t,t_0)$ evaluated over one period of the driving:
\begin{equation}
 U(T+t_0,t_0)\Psi_{}(x,t_0)=e^{-i\epsilon_{{}} T}\Psi_{{}}(x,t_0).
 \label{eq:time_evolution}
\end{equation}
By successively solving the above eigenvalue equation for different $t_0$ ranging from $t_0=0$ to $t_0=T$, one obtains the full time dependence of $\Psi_{{}}(x,t)$.
This is done numerically, whereby the major task is to actually compute a matrix representation of the time evolution operator $U(T+t_0,t_0)$. 
An efficient scheme which allows for the extraction of the Floquet modes by means of Eq. (\ref{eq:time_evolution}) is explained in detail in Ref.\,\cite{wulf_symmetries_2014, wulf_site-selective_2015}.

%%%%%%%%%%%%%%%%%%%%%%%%%%%%%%%%%%%%%%%%%%%%%%%%%
\begin{figure}[t!]
\centering
\includegraphics[width=1\columnwidth]{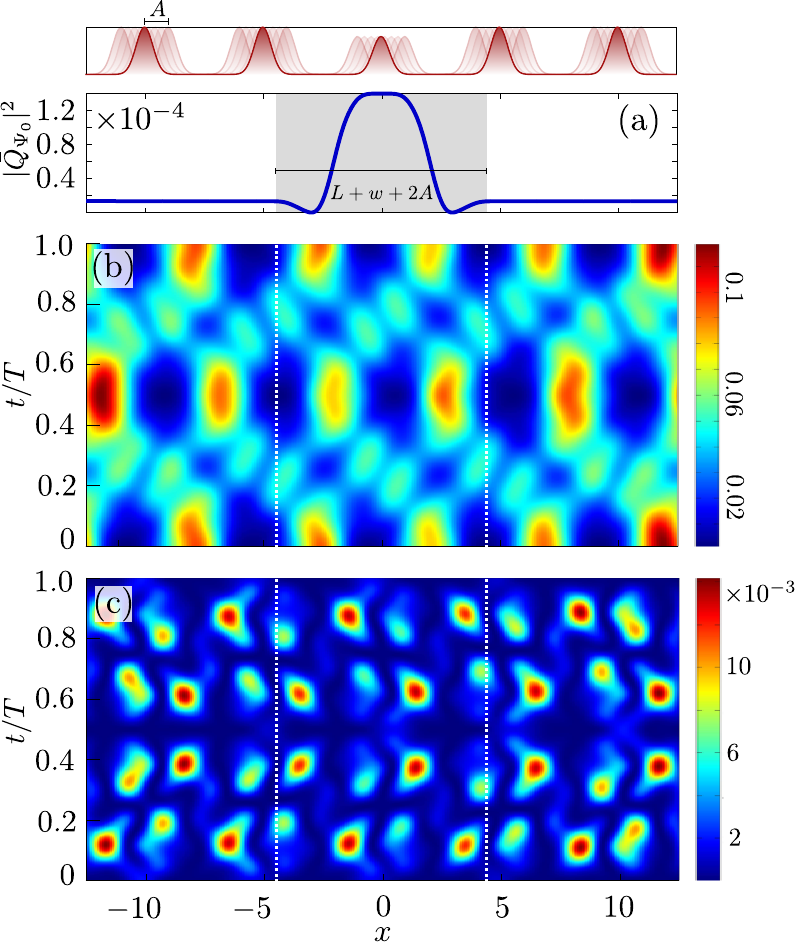}
\caption{\label{fig:driven} 
(a) Magnitude squared of the locally invariant current $\bar {{Q}}_{\Psi_0}$ (again shifted by $+L/2$ in $x$) of the Floquet mode $\Psi_0$ for the setup in Fig.\,\ref{fig:static}\,(a1), but now oscillating with amplitude $A=1$ and frequency $\omega =0.5$. 
(b) Density plot, $|\Psi_0|^2$, of the Floquet mode used in (a). (c) Integral kernel $|{{Q}}_{\Psi_0}(x,t)|^2$ (cf. Eq. (\ref{eq:Qdynamical})) for one period of the driving.}
\end{figure}
%%%%%%%%%%%%%%%%%%%%%%%%%%%%%%%%%%%%%%%%%%%%%%%%%

We now apply this technique to the previous case of $N=5$ Gaussian barriers with a defect on the central barrier, with the whole lattice laterally driven according to Eq.\,(\ref{eq:driving}) with $A=1.0$ and $\omega = 0.5$
and for the Floquet mode $\Psi_0(x,t)$ with the lowest time averaged energy (see Fig.\,\ref{fig:driven} (b) for a density plot of $\Psi_0(x,t)$). 
At any instance in time, the spatial symmetry under translation by $L$ is thus locally broken around the central barrier.
In Fig. \ref{fig:driven} (a) we show that $|\bar {Q}(x)|^2$ is indeed spatially constant except around the defect position. 
Note that, over one period of driving, the defect has an increased effective width $w+2A$ due to the lateral shaking, so that the region of deviation from $\bar {{Q}}'=0$ is increased by $2A$ compared to the static case (cf. Fig.\,\ref{fig:static}\,(b1)).
Figure \ref{fig:driven} (c) also shows the absolute square $|{{Q}}(x,t)|^2$ of the integrand in Eq. (\ref{eq:Qdynamical}), confirming that indeed only the time-averaged current is invariant.
In summary, even in a regime where it is difficult to deduce symmetries of the potential---or local breaking thereof---directly from the wavefunction, 
this is accomplished reliably with the introduced dynamical local invariants.

\subsection{Local symmetry based convergence measure}
\label{sec:convergence}

So far we have seen how the local symmetries of a system are encoded in its stationary or quasistationary states and can be extracted by calculating the two-point currents ${{Q}}$ or $\bar Q$ respectively. 
In this sense, we obtain structural information on the wavefunction in situations where global symmetries are absent but local symmetries are present as the resulting ${{Q}}$'s have to be constant within the domain of LS. 
When confronted with the task to numerically solve the Schr\"odinger equation for its stationary or quasistationary states in the presence of global symmetries, it is a usual procedure to 
directly build in the symmetries into the computational scheme (e.g. by making sure that the basis functions respect the restrictions imposed by the symmetry). %, or in case that this is not possible to
%check their precision through the fulfillment of symmetry induced restrictions.
% For example, a standard criterion in the non degenerate time-independent case is the symmetry of the density, $|\Psi(x)|^2 = |\Psi(-x)|^2$ for $V(x)=V(-x)$.
However, this can only be done if the symmetry is global.
We now outline a strategy of exploiting the additional information gained by the LS analysis in order to check the convergence of the numerical solution of the Schr\"odinger equation in systems with broken global symmetry but retained local symmetry.

\begin{figure}[t!]
\centering
\includegraphics[width=.8\columnwidth]{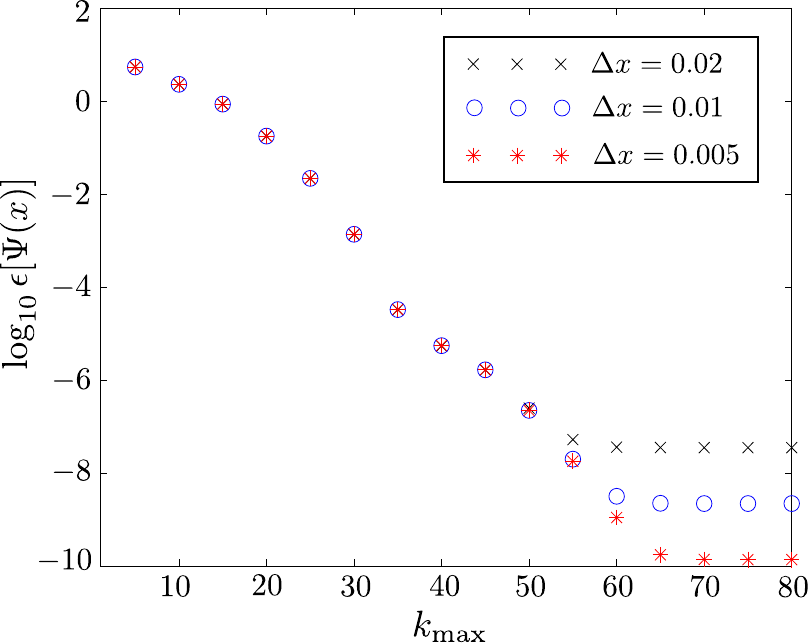}
\caption{\label{fig:convergence} 
Convergence measure $\epsilon_\Psi$ on a logarithmic scale for the ground state of the setup in Fig.\,\ref{fig:static}\,(a1) in dependence of $k_{\text{max}}$ (with the plane wave basis size being $2k_{\text{max}}+1$) for three different lattice spacings $\Delta x$ of the numerical grid on which $\Psi(x)$ is represented.} 
\end{figure}

% For simplicity, we assume that stationary solutions of a static potential are computed, with the application for a quasistationary solution (Floquet mode) of a periodically driven potential being analogous. 
% Let the potential respect a LS within some domain $\mathcal{D}$, i.\,e. $V(x)=V(\bar{x})$ for all $x, \bar{x} \in \mathcal{D}$.
The idea is here to define a quantity which `measures' the deviations from a constant current ${{Q}}$ or $\bar Q$ within the LS domain. While there is no unique way to define such a quantity, a straightforward one is given by the current's derivative 
integrated over $\mathcal{D}$,
\begin{equation}
 \epsilon_\Psi \equiv D^2 \frac{\int_{\mathcal{D}}\,dx |{{Q}}_\Psi'|^2}{\int_{\mathcal{D}}\,dx |{{Q}}_\Psi|^2}.
\end{equation}
where $Q_\Psi$ has to be replaced by $\bar Q_\Psi$ for a periodically time-dependent setting and
multiplication by the size $D$ of $\mathcal{D}$ as well as dividing by the average of $|{{Q}}_\Psi(x)|^2$ are introduced merely to render $\epsilon_\Psi$ dimensionless.
This convergence indicator should ideally be zero, so that, by construction, any deviation from $\epsilon_\Psi=0$ indicates a numerical error in the computed wavefunction.
As an illustrative example, we return to the static setup of 5 Gaussian barriers with an attractive central defect (see Figs.\,\ref{fig:static}\,(a1,b1,c1)). 
Figure \ref{fig:convergence} shows the logarithmic dependence of $\epsilon_{\Psi}$ for the system's ground state on the maximal number $2k_{\text{max}}+1$ of plane waves used as a basis (cf. Sec.\,\ref{sec:static_lattice}). 
Since the proposed convergence measure is a functional of the wavefunction, it does not only provide information on the accuracy of the truncated matrix representation of the Hamiltonian, but also on the accuracy with which the wavefunction is represented on a numerical grid. 
Hence, it clearly indicates at which point an increasing number of basis states does not lead to further agreement of the wavefunction with the restrictions imposed on it by the invariance of ${{Q}}_\Psi(x)$, as this is limited by the finite spacing $\Delta x$ of the grid.

%%%%%%%%%%%%%%%%%%%%%%%%%%%%%%%%%%%%%%%%%%%%%%%%%
%%%%%%%%%%%%%%%%%%%%%%%%%%%%%%%%%%%%%%%%%%%%%%%%%

\section{Relation of invariants to probability current conservation and time reversal invariance}
\label{sec:transfer_matrices}

As the results shown in this manuscript hinge on the spatial constancy of the two point currents ${{Q}}$ and $\bar {{Q}}$, it would certainly be desirable 
to gain some physical insight into these currents and in particular to link their invariance to known physical principles.
For the static case, we show how this can be done in the following by providing a link of the spatial invariance of $Q$ to probability current conservation (PCC) and time reversal invariance (TRI) in 1D time-independent scattering. 
From the fact that PCC and TRI do not hold in general for periodically driven systems, it is clear that a straightforward generalization of the results presented in this section is not possible (see also discussion at the end of this section).

We consider the static version of the potential in Eq.\,(\ref{eq:lattice_pot}) with arbitrary scatterer potentials $V_n(x)$ of width $w<L$. 
Since $V(x)=0$ between the scatterers, the wavefunction can be expanded in these regions in terms of counter-propagating plane waves as $\Psi_n(x) = F_n e^{ikx} + G_n e^{-ikx}$ for $x \in (X_{n-1}+w/2,X_n-w/2)$ at a given energy $E=k^2 / 2$.
The wavefunction amplitudes within adjacent potential free regions are then related via 
\begin{equation}
\mathbf{\Psi}_{n+1} = \mathbf{{M}}_{n} \mathbf{\Psi}_{n},  \quad
\mathbf{\Psi}_{n} \equiv \left( \begin{array}{c} F_{n} \\ G_{n} \end{array} \right),
\label{eq5}
\end{equation}
where the transfer matrix $\mathbf{{M}}_{n}$ propagates the plane wave amplitudes from left to right across the scatterer centered at $X_{n}$. 
$\mathbf{{M}}_{n}$ is therefore given by the transfer matrix $\mathbf{P}_n$ of the localized potential $V_n$ at the origin, shifted to the position $X_n = nL$ by a diagonal phase shift matrix $\mathbf{K}_{s=X_n}$ (see e.g.\cite{mello2004quantum}):
\begin{equation}
\mathbf{M}_{n} = \mathbf{K}_{nL}^* \mathbf{P}_{n} \mathbf{K}_{nL}, ~~~
\mathbf{K}_{s} = \begin{pmatrix} e^{iks} & 0 \\ 0 & e^{-iks} \end{pmatrix}.
\label{M_n}
\end{equation}
The total transfer matrix through the whole scattering region is then the ordered product $\mathbf{M} = \mathbf{M}_{N-1}\mathbf{M}_{N-2} \cdots \mathbf{M}_0$ of those transfer matrices.

Let us now assume that the chain of scatterers possesses a local translation or inversion symmetry over a domain $\mathcal{D}$ with $N_\mathcal{D}$ scatterers starting from the $n = d$-th one, that is, we have $V(x) = V(\bar{x})$ for $x,\bar{x} \in \mathcal{D} = (X_d - w/2, X_{d+N_\mathcal{D}-1})$ with $\bar{x}= x+L$ (translation) or $\bar{x}= -x+2\alpha$ (inversion), respectively.
The main point of this section is then to show that the local invariance of ${{Q}}$ and $Q^\text{c}$ in $\mathcal{D}$ can be obtained from the LS of the potential together with two very general properties of $\mathbf{M}_{n}$ arising from (a) PCC and (b) TRI in 1D Hermitian, time independent quantum mechanics. 
In terms of the transfer matrices, these two conditions imply (see e.g. \cite{mello2004quantum}) that
\begin{subequations}  \label{conditions}
 \begin{align} 
\label{eq:eta}
 \mathbf{M}^\dagger_{n} \boldsymbol{\eta} \mathbf{M}_{n} &= \boldsymbol{\eta} \equiv\begin{pmatrix} 1 & 0 \\ 0 & -1 \end{pmatrix},\ (\text{PCC}) \\
\label{eq:TRI}
 \mathbf{M}_n^{-1} \boldsymbol{\tau} \mathbf{M}_n^* &= \boldsymbol{\tau} \equiv \begin{pmatrix} 0  & 1 \\ 1 & 0 \end{pmatrix},\ (\text{TRI})
 \end{align}
\end{subequations}
respectively. % , with $\mathbf{M}^\dagger_{n}\equiv (\mathbf{M}^*_{n})^T$. 
The combination of PCC with TRI in turn requires the transfer matrices to be unimodular: $\text{det}(\mathbf{M}_{n})=1$. 
In the following, we demonstrate the relation of local invariance of ${{Q}}$ to the above conditions (a) and (b) separately for translation and inversion symmetry.

In the case of a local translation by $L$, the two-point current ${{Q}}$ in the potential-free region $(X_{n-1}+w/2,X_{n}-w/2)$ between scatterers $n-1$ and $n$ reads ${{Q}}_n = k \left( F^*_{n}F_{n+1}e^{ikL} - G^*_{n}G_{n+1}e^{-ikL} \right)$, which can be written as the scalar product
\begin{equation} \label{eq:Qn_trans}
{{Q}}_n = k\mathbf{\Psi}_n^\dagger \boldsymbol{\eta} \mathbf{K}_L \mathbf{M}_{n}\mathbf{\Psi}_n.
\end{equation}
If the lattice potential is translation invariant within a domain $\mathcal{D}$, then we have $V_n=V_{n+1}$ and hence
\begin{equation} \label{eq:LT_condition}
 \mathbf{M}_{n+1}= \mathbf{K}_L^* \mathbf{M}_{n} \mathbf{K}_L
\end{equation}
for the corresponding transfer matrices.
Using the form of ${{Q}}_n$ in Eq.\,(\ref{eq:Qn_trans}), it is straightforward to show that Eq.\,(\ref{eq:LT_condition}) leads indeed to ${{Q}}_n={{Q}}_{n+1}$ 
if we assume additionally the condition imposed on the transfer matrix by PCC (Eq.\,\ref{eq:eta}).
In this sense, one may interpret the invariance of the two-point current ${{Q}}$ within a domain of local translational symmetry as a direct consequence of the global invariance of the probability current.

In the case of a local inversion through $\alpha$ within $\mathcal{D}$, the two-point currents become ${{Q}}_n = k \left( F^*_{n}G_{\bar{n}+1}e^{-2ik\alpha} - G^*_{n}F_{\bar{n}+1}e^{2ik\alpha} \right)$ between scatterers $n-1$ and $n$, compactly written as
\begin{equation} \label{eq:Qn_inv}
{{Q}}_n  = k \mathbf{\Psi}_n^\dagger \boldsymbol{\zeta} \mathbf{K}_{2\alpha} \mathbf{M}_{\bar{n}} \mathbf{\Psi}_{\bar{n}},
\end{equation}
where we recall that the interval $(X_{n-1},X_{n})$ is mapped to $(X_{\bar{n}},X_{\bar{n}+1})$ with $\bar{n} = 2\alpha/L - n$ in the present enumeration of scatterers (assuming here an odd number of scatterers in $\mathcal{D}$ and $n>0$ without loss of generality).
On the other hand, the local inversion symmetry requires that
\begin{equation} \label{eq:LP_condition}
 \mathbf{M}_{\bar{n}}^{-1} = \mathbf{K}_{2\alpha}^* \mathbf{M}_n^* \mathbf{K}_{2\alpha}
\end{equation}
for the transfer matrix of scatterer $n$, which can be shown by considering the time-reversed propagation with $X_n$ shifted by $2\alpha$ to $X_{-\bar{n}}$.
Under this condition, we can now show straightforwardly by using Eq.\,(\ref{eq:Qn_inv}) that ${{Q}}_n={{Q}}_{n+1}$ if 
\begin{equation} \label{eq:zeta}
 \mathbf{M}^\dagger_{n} \boldsymbol{\zeta} \,\mathbf{M}_{\bar{n}}^* = \boldsymbol{\zeta} \equiv \boldsymbol{\eta} \boldsymbol{\tau}  = \begin{pmatrix} 0 & 1 \\ -1 & 0 \end{pmatrix},
\end{equation}
which in turn follows from Eqs.\,(\ref{eq:eta}) and (\ref{eq:TRI}) combined.
Thus, in contrast to the translation symmetry case, the invariance of ${{Q}}$ in a local inversion symmetry domain requires TRI in addition to PCC.

In complete analogy to the above, the invariance of the complementary, time-dependent two-point quantity $Q^\text{c}$ (with $\mathbf{\Psi}_n^\top$ replacing $\mathbf{\Psi}_n^\dagger$ in Eqs.\,(\ref{eq:Qn_trans}),\,(\ref{eq:Qn_inv}) for the plane wave representation) can also be linked to PCC and/or TRI.
Opposite to the case of ${{Q}}$ though, we now obtain $Q^\text{c}_n = Q^\text{c}_{n+1}$ in an inversion LS domain from PCC alone and in a translation LS domain from combined PCC and TRI.
Note that, whereas PCC and/or TRI are sufficient for locally invariant ${{Q}}$ and $Q^\text{c}$ in a LS domain, they are not \textit{necessary} conditions.
Indeed, ${{Q}}$ and $Q^\text{c}$ have been shown to be spatially constant also for non hermitian static potentials with antisymmetric \cite{kalozoumis_systematic_2014} and symmetric \cite{kalozoumis_invariant_2015} imaginary part, respectively, for a given symmetry transformation.
For such potentials PCC and TRI generally do not apply, but the transfer matrix of a localized scatterer still retains its unimodularity ($\text{det}(\mathbf{M})=1$).
Similarly, the driven systems considered in the previous sections, for which sideband transfer matrices can in fact be defined \cite{Yakubo1998}, do give rise to (generalized) invariant currents $\bar {{Q}}$ in spite of not supporting PPC or TRI generally. 
An interesting prospect for future work would thus be the connection of the symmetry-induced two-point invariants to fundamental conservation and invariance principles in non hermitian and/or driven wave-mechanical systems.

\section{Conclusions}
\label{sec:conclusion}

The theory of invariants of broken discrete symmetries \cite{kalozoumis_invariants_2014} has been generalized to periodically driven systems by means of Floquet theory, and subsequently applied to superlattice setups.
In particular, we derived a dynamical two-point current $\bar {{Q}}(x,\bar{x})$ which is spatially constant for quasistationary states of arbitrary time-periodic potentials within domains where 
the potential is invariant under the transformation $x \rightarrow \bar x$. 
With the theory applying equally to local symmetry under translation or inversion and for arbitrary boundary conditions, we focused here on the case of translation symmetries in lattice potentials locally deformed by defects.
It was thereby demonstrated that the positions of such defects or any deviation from a given symmetry can be extracted reliably from the stationary or quasistationary states for static or driven systems by use of the two-point invariants.  
As a potential application, we also argued how the restriction of a constant two-point current on the wavefunctions can be exploited in order to test the numerical convergence of a computed solution for systems with broken global symmetry.
Finally, using a transfer matrix formalism we linked the local invariance of the symmetry-induced currents to two generically present properties of Hermitian 1D quantum mechanics: time reversal invariance and probability current conservation.
As our results do not dependent on the specific form of the potential, we believe that they are of relevance for a variety of physical realizations of static or driven lattice systems such as cold atoms immersed in optical lattices \cite{bloch_ultracold_2005, salger_directed_2009}, radiated semiconductors \cite{glazov_high_2014}, or photonic devices \cite{rechtsman_photonic_2013}.

%%%%%%%%%%%%%%%%%%%%%%%%  APPENDIX
%%%%%%%%%%%%%%%%%%%%%%%%              APPENDIX
\appendix
\section{Nonlocal invariants for a point scatterer array}
\label{A1}

We here investigate the two-point current ${{Q}}(x)$ for an array of $N$ delta potentials located at positions $X_n\equiv nL$ for $n=0,1,...,N-1$ and containing a single defect. 
We thus consider a potential of the type as given in Eq. (\ref{eq:lattice_pot}) and set
\begin{equation}
  V_n(x) = \Lambda_n \delta(x),
\end{equation}
with site-dependent potential strengths $\Lambda_n$. 
% The transfer matrices $\mathbf{P}_n$ (cf. Eq. (\ref{M_n})) then take the simple form \cite{wave_prop}:
% \begin{equation}
% \mathbf{P}_n =
% \begin{pmatrix} 1+a_n & a_n \\ a^*_n & 1+a^*_n \end{pmatrix}
% \label{P}
% \end{equation}
% with $a_n(k) = \Lambda_n/ik$.
We choose odd $N$ and set $\Lambda_n = \Lambda$ except for the central peak at site $c=(N-1)/2$ with $\Lambda_c\neq \Lambda$.
Then the system is for $0<x<(N-1)L$ translationally symmetric with $V(x)=V(x+L)$ except for the defect located at $X_c$. 
Let us now calculate the two-point current ${{Q}}$ for the assumed translational symmetry of $V(x)=V(x+L)$. 
Since the entire lattice does not respect the symmetry of a translation by $L$ (because of the impurity), 
the two-point current will, in general, not be invariant and thus becomes a function of $x$, ${{Q}}={{Q}}(x)$. 
However, away from the impurity, i.\,e. anywhere except in the domain $x\in (X_{c-1},X_c)$, the translational symmetry of $V(x)=V(x+L)$ holds and so ${{Q}}(x)$ is expected to be constant. 
% In terms of the domain-wise notation of ${{Q}}(x)$ this means that ${{Q}}_n={{Q}}_{n+1}$ for $n<c$ and equally for $n>c$.
Consequently, ${{Q}}(x)$ takes the form of a piecewise constant function:
\begin{equation}
    {{Q}}(x)=
    \begin{cases}
      {{Q}}_{\mathcal{L}}, & \text{if}\ x<X_c \\
      {{Q}}_{\mathcal{C}}, & \text{if}\ X_{c-1}<x<X_c \\
      {{Q}}_{\mathcal{R}}, & \text{if}\ x>X_c
    \end{cases}.
    \label{analytic_cases}
\end{equation}

%%%%%%%%%%%%%%%%%%%%%%%%%%%%%%%%%%%%%%%%%%%%%%%%%
\begin{figure}[t!]
\centering
\includegraphics[width=.8\columnwidth]{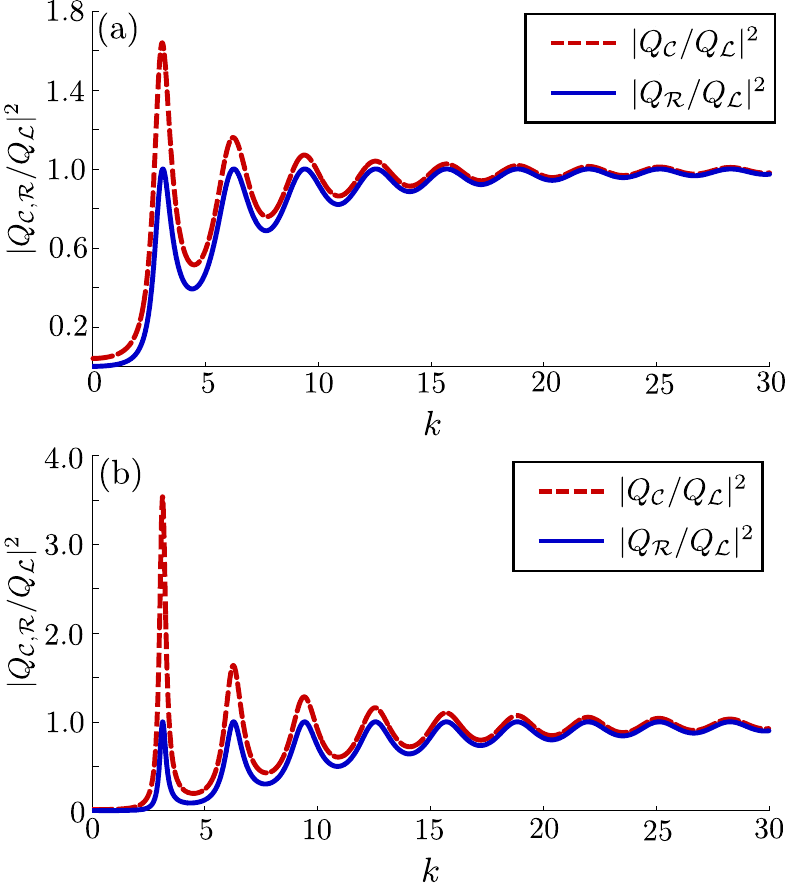}
\caption{\label{fig:delta} 
Ratios of the absolute squares of the three different values ${{Q}}_{\mathcal{L}}$, ${{Q}}_{\mathcal{C}}$ and ${{Q}}_{\mathcal{R}}$ that the current ${{Q}}(x)$ takes 
for a single point scatterer located at $X_c$ (see Eq. (\ref{analytic_cases})) as functions of the momentum $k$ of the incoming wave. 
Shown are the analytic results stated in Eq. (\ref{eq:delta_analytic}) for two different strengths (a) $\Lambda_c=2.5$ and (b) $\Lambda_c=5$ of the delta potential.} 
\end{figure}
%%%%%%%%%%%%%%%%%%%%%%%%%%%%%%%%%%%%%%%%%%%%%%%%%

In order to present reasonably simple and thereby insightful expressions for ${{Q}}_{\mathcal{L}}$, ${{Q}}_{\mathcal{C}}$, and ${{Q}}_{\mathcal{R}}$, we may simplify the setup even further:
Without loss of generality (for the present LS analysis), we set $\Lambda = 0$ and end up with a single point scatterer at $x=X_c$.
% Thus, the transfer matrices with $n\neq c$ reduce to unit matrices, $\mathbf{M}_{n\neq c} = \mathbf{I}$, and the total transfer matrix becomes $\mathbf{M} = \mathbf{M}_{c}$. 
If we then consider a plane wave with amplitude $F_0$ coming in from the left, the invariant current ${{Q}}(x)$ (cf. Eq.\, (\ref{analytic_cases}) can be calculated as
\begin{equation}
\begin{aligned} 
&{{Q}}_{\mathcal{L}} &&= &&A_{kL} \left(1 - \frac{|a_c|^2}{1+|a_c|^2}e^{-2ikL} \right) \\
&{{Q}}_{\mathcal{C}} &&= &&A_{kL}\frac{1}{1+a_c^*}\\
&{{Q}}_{\mathcal{R}} &&= &&A_{kL} \frac{1}{1+|a_c|^2}
\label{eq:delta_analytic}
\end{aligned}
\end{equation}
with $A_{kL} \equiv k |F_0|^2 e^{ikL}$ and $a_c = \Lambda_c/ik$.
% Note that, in the case of multiple peaks with the same $\Lambda_n=\Lambda_c$ instead of a single peak, we would obtain three plateaus of different ${{Q}}_n$, but their analytic expressions would not be as simple as above.

A major advantage of such a minimal setup is that we can readily check some limiting cases.
Let us therefore study ${{Q}}(x)$ for the single scatterer in the two limiting cases of small and large potential strength. 
For a vanishing potential compared to the energy of the incoming wave, $\Lambda_c/k \rightarrow 0$, we get $a_c \rightarrow 0$ and thus a common value 
${{Q}}_{\mathcal{L}} = {{Q}}_{\mathcal{C}} = {{Q}}_{\mathcal{R}} = A_{kL}$. 
Note that this is expected since this limit corresponds to a free particle propagation which has a global (and in fact continuous) translation invariance and thus is associated with a globally constant current ${{Q}}(x)$.
In the opposite limit, $\Lambda_c/k \rightarrow \infty$, the transmission through the potential becomes zero.
Thus, the wavefunction at $x>X_c$ vanishes, and thereby ${{Q}}_{\mathcal{R}}$ and ${{Q}}_{\mathcal{C}}$ vanish too.
For this case of total reflection, in the left region ($x<X_{c}$) we have a sinusoidal standing wave and so ${{Q}}_{\mathcal{L}} = 2ik|F_0|^2\sin(kL)$ (this can be calculated straightforwardly from Eq.\,(\ref{eq:Qn_trans}).
The full energy dependence of the absolute squares of the ratios ${{Q}}_{\mathcal{R}} / {{Q}}_{\mathcal{L}}$ and ${{Q}}_{\mathcal{C}} / {{Q}}_{\mathcal{L}}$ 
are shown in Fig. \ref{fig:delta} for different strengths $\Lambda_c$ of the potential. 

Interestingly, we see that although $|{{Q}}_{\mathcal{C}} / {{Q}}_{\mathcal{L}}|^2=1$ or $|{{Q}}_{\mathcal{R}} / {{Q}}_{\mathcal{L}}|^2=1$ for some momenta $k$, we observe 
no momentum where ${{Q}}(x)$ is globally constant, i.\,e. where $ {{Q}}_{\mathcal{R}} = {{Q}}_{\mathcal{C}}= {{Q}}_{\mathcal{L}}$.
Hence, we conclude that the local breaking of translational symmetry by the single point scatterer is always accompanied by a deviation from a globally constant ${{Q}}$.

\bibliography{loc_sym_sta_dri_lat}{}

\end{document}